\documentclass[a4paper,11pt]{article}
\usepackage{graphicx}% Include figure files
\usepackage{dcolumn}% Align table columns on decimal point
\usepackage{bm}% bold math
\usepackage{epsfig,amsmath}
\usepackage[usenames,dvipsnames]{color}
\usepackage[pagebackref=false, colorlinks=true]{hyperref}
\usepackage{textcomp}
\usepackage{subfigure}
\definecolor{redish}{rgb}{0.7,0.2,0.0}  % color defined in (r=red,g=green,b=blue) model
\definecolor{bluish}{rgb}{0.2,0.5,0.8}

\hypersetup{linkcolor=redish,          % color of internal links
                  citecolor=blue,        % color of links to bibliography
                  filecolor=magenta,      % color of file links
                  urlcolor=bluish}          % color of the url

\DeclareFontFamily{U}{rsfs}{}         % Formal Script            %
\DeclareFontShape{U}{rsfs}{m}{n}{<5> rsfs5 <6><7> rsfs7          %
  <8><9><10><10.95><12><14.4><17.28><20.74><24.88> rsfs10}{}     %
\DeclareMathAlphabet{\mathfs}{U}{rsfs}{m}{n}                     %

\newcommand{\ba}{\nopagebreak[3]\begin{eqnarray}}
\newcommand{\ea}{\end{eqnarray}}

\newcommand{\f}{\frac}

\def \O{\Omega}

\def \d{\Delta}

\def \th{\theta}

\title{Anomalous Lense-Thirring precession in 
Kerr-Taub-NUT spacetimes}
\author{Chandrachur Chakraborty\footnote{chandrachur.chakraborty@saha.ac.in} 
 \\ {\it Saha Institute of Nuclear
Physics,Kolkata-700064,India}}

\date{}

\begin{document}

\maketitle

\begin{abstract}
Exact Lense-Thirring (LT) precession in Kerr-Taub-NUT
spacetime is reviewed.
It is shown that the LT precession does not obey the 
general inverse cube law of distance at strong gravity regime in
Kerr-Taub-NUT spacetime. Rather, it becomes maximum just near
the horizon, falls sharply and becomes zero near the horizon.
The precession rate increases again and after that it falls
obeying the general inverse cube law of distance.
This anomaly is maximum at the polar region of
this spacetime and it vanishes after crossing 
a certain `critical' angle towards equator from pole.
We highlight that this particular `anomaly' also arises
in the LT effect at the interior spacetime of the pulsars and
such a signature could be used to identify a role of Taub-NUT solutions
in the astrophysical observations or equivalently, a signature of the existence of NUT
charge in the pulsars. In addition, we show that if the Kerr-Taub-NUT 
spacetime rotates with the angular momentum $J=Mn$ (Mass$\times$Dual Mass),
inner horizon goes to at $r=0$ and only {\it event horizon}
exists at the distance $r=2M$. 
\end{abstract}

{\bf Keywords} Lense-Thirring precession . Strong Gravity . 
NUT Charge . Taub-NUT spacetime . Pulsar

\section{Introduction}
The Lense-Thirring (LT) precession\cite{lt} is a important phenomena
in General relativity as well as in Relativistic astrophysics.
In this phenomena the locally inertial frames are dragged 
along the rotating spacetime due to the 
angular momentum of the stationary spacetime. LT precession rate is 
proportional to the curvature as well as 
the angular velocity of the 
rotating spacetime. Thus, the effect will be
larger for massive and rapidly rotating spacetime around which
the curvature effect is maximum. 
It is perhaps Majumdar and myself \cite{cm} first to
motivate investigation on the LT precession in strong
gravity situation. Without making any preliminary
assumption we have derived the exact LT precession
rate of a gyroscope in the Kerr and Kerr-Taub-NUT (KTN) spacetimes.
Now, it can easily be obtained the frame-dragging rate of
a gyroscope\cite{schiff} just outside a Kerr spacetime\cite{rk}
as well as a KTN spacetime\cite{taub, nut}.
Kerr and KTN spacetimes both are the vacuum solutions of Einstein
equation. Kerr spacetime has two parameters: mass and Kerr
parameter (angular momentum per unit mass) but
there are three parameters to describe the KTN spacetime.
The parameters are: mass, Kerr 
parameter and NUT parameter. If the NUT parameter vanishes
the KTN spacetime reduces to the Kerr spacetime and if the Kerr 
parameter vanishes the KTN spacetime reduces to the Taub-NUT
spacetime. In the absence of the NUT parameter, the Taub-NUT 
spacetime reduces to pure Schwarzschild spacetime
which is non-rotating. The Kerr spacetime is very
well known to us and it is also physically reliable. 
We can describe the exterior geometry of 
many rotating astrophysical objects by the Kerr spacetime
only in the approximation when the multipole momenta of the rotating matter are negligible.
Otherwise, the metric receives corrections from higher gravitational multipole moments \cite{sg}.
Thus, in the general sense the Kerr spacetime is astrophysically relevant
but the KTN spacetime is quite different than the Kerr geometry. As it holds an
additional parameter (NUT), this spacetime does not
physically relevant till now.

 Lynden-Bell and Nouri-Zonoz \cite{lnbl}
are the first to motivate investigation on the
observational possibilities for NUT charges
or (gravito)magnetic monopoles. They have claimed
that the signatures of such spacetime might be
found in the spectra of supernovae, quasars, or active 
galactic nuclei. It has also been recently brought into 
focus by Kagramanova et. al \cite{kag} by a detail
and careful analysis of geodesics in the Taub-NUT spacetime.
A rigorous analysis in extremal and non-extremal KTN spacetimes
for timelike and spacelike geodesics has already been done by myself \cite{cc}.
It should be noted that the (gravito)magnetic monopole
spacetime with angular momentum (basically the KTN spacetime) admits relativistic 
thin accretion disks of a black hole in a galaxy 
or quasars \cite{liu}. The accretion disks are basically formed 
just near the above mentioned astrophysical objects.
In this sense the accretion phenomena takes place in a very strong 
gravity regime where the frame-dragging effect
is expected to be very high. Thus the frame-dragging 
effect should have greater impact
on accretion disk phenomena. This provides us a strong 
motivation for studying the LT precession or frame-dragging
effect in the KTN spacetime in more detail because it will affect the accretion 
in such spacetimes from massive stars, and might offer 
novel observational prospects.

The KTN spacetime 
is a stationary and axisymmetric vacuum solution of Einstein
equation. This spacetime consists of the Kerr and 
NUT parameters. The Kerr parameter is responsible for
the rotation of the spacetime. In general sense the NUT 
charge should not be responsible explicitly for the rotation
of the spacetime but implicitly this NUT charge can add a ``rotational
sense'' in a non-rotating spacetime.
The NUT charge is also called as `dual mass' whose properties
 have been investigated in detail by Ramaswamy and Sen \cite{sen2}.
They also called the NUT parameter as the ``angular momentum monopole'' \cite{rs}
which is quite sound in this sense that it can give a ``rotational sense''
of the Taub-NUT spacetime even when the Kerr parameter vanishes.
In this regards, though the Kerr parameter vanishes
in the KTN spacetime, the Taub-NUT spacetime retains
the rotational sense due to the NUT parameter. Due to the 
presence of the NUT parameter the spacetime still remains
stationary and violates the time reflection symmetry.
Time reflection changes the direction of rotation
and thus does not restore one to the original configuration \cite{rw}.
Thus, the failure of the hypersurface orthogonality
(it also means that the spacetime preserves the time translation symmetry
but violates the time reflection symmetry) 
condition implies that the neighbouring orbits of 
$\xi^a$ (the timelike Killing vector which must exist
in any stationary spacetime) ``twist'' around each other. 
In the Kerr spacetime, the presence of the Kerr parameter 
makes the spacetime stationary instead of static.
Similarly, in the case of the Taub-NUT spacetime the NUT
parameter compels the spacetime stationary instead of
static. Thus, the Kerr and NUT parameters
both are responsible to make the spacetime in rotation.
Thus, it is needless to say that the KTN spacetime 
must be stationary. 
 
The strong gravity LT precession in KTN spacetime has
already been highlighted in \cite{cm} by Majumdar and myself
but we could not studied it in detail. 
Though our target was to investigate LT
precession in Kerr and KTN spacetimes, it had taken a 
turn into the investigation of LT precession in
Taub-NUT spacetime which was really very interesting in
that situation. We were busy to shown that the LT precession
could not vanish even in {\it non-rotating} (as Kerr parameter vanishes)
Taub-NUT spacetime. Later, Modak, Bandyopadhyay and myself \cite{cmb}
have discovered that frame-dragging curves are
not smooth along the equator and its surroundings 
inside a rotating neutron star. Rather, the frame-dragging effect shows an 
interesting {\it anomaly} along the equator inside the pulsars. The frame-dragging rate is
maximum at the center and decreases initially away from the
center, tends to zero (not exactly zero but very small) before
the surface of the neutron star, rises again and finally approaches
small value on the surface as well as outside of the pulsars.
We think that this may not be the only case where 
we see this {\it anomaly}. After that we start to hunt for
this type of feature in other spacetimes which are the vacuum
solutions of Einstein equation and we get the almost
similar {\it anomaly} in the KTN spacetime (we note that there are many
differences between the KTN spacetime and the spacetime 
of a rotating neutron star; they are not same).
Previously, the strong gravity LT precession in Pleba\'nski-Demia\'nski (PD)
spacetimes (most general axisymmetric and stationary spacetime
till now) has been investigated by Pradhan and myself \cite{cp}.
But our close observation says 
that due to the presence of the NUT charge this anomaly 
in the frame-dragging can also arise 
in the PD spacetime. In this present paper, we are 
now investigating only for LT precession in KTN
spacetime as it may be astrophysically sound in near future.

The paper is organized accordingly as follows: in section 2,
we review the LT precession in KTN spacetime. We also discuss
a very special case of the KTN spacetime in a subsection
of the section 2. We discuss our result in
section 3 and finally, we conclude in section 4.

\section{Lense-Thirring precession in Kerr-Taub-NUT spacetime}
The KTN spacetime is a geometrically stationary and
axisymmetric vacuum solution of Einstein equation.
This spacetime consists mainly three parameters: mass $(M)$,
angular momentum $(J)$ per unit mass or Kerr parameter $(a=J/M)$ 
and NUT charge $(n)$ or dual mass.  
 The metric of the KTN spacetime can be written as \cite{ml}
\begin{equation}
ds^2=-\f{\d}{p^2}(dt-A d\phi)^2+\f{p^2}{\d}dr^2+p^2 d\th^2
+\f{1}{p^2}\sin^2\th(adt-Bd\phi)^2
\label{lnelmnt}
\end{equation}
With 
\begin{eqnarray}\nonumber
\d&=&r^2-2Mr+a^2-n^2,  p^2=r^2+(n+a\cos\th)^2,
\\
A&=&a \sin^2\th-2n\cos\th, B=r^2+a^2+n^2.
\end{eqnarray}

The exact LT precession rate in the KTN spacetime
is (Eq.(20) of \cite{cm})
\begin{eqnarray}
\vec{\O}_{LT}=\f{\sqrt{\d}}{p}\left[\f{a \cos\th}{\rho^2-2Mr-n^2}
-\f{a \cos\th+n}{p^2}\right]\hat{r}
+\f{a \sin\th}{p} \left[\f{r-M}{\rho^2-2Mr-n^2}
-\f{r}{p^2}\right]\hat{\th}
\label{kt}
\end{eqnarray}
where, $\rho^2=r^2+a^2\cos^2\th$.
The modulus of the above LT precession rate is
\begin{eqnarray}
\O_{LT}=|\vec{\O}_{LT}|=\f{1}{p}\left[\d\left(\f{a \cos\th}{\rho^2-2Mr-n^2}
-\f{a \cos\th+n}{p^2}\right)^2+a^2\sin^2\th \left(\f{r-M}{\rho^2-2Mr-n^2}
-\f{r}{p^2}\right)^2\right]^{\f{1}{2}}
\label{ktmod}
\end{eqnarray}
It could be easily seen that the above equation is valid
only in timelike region, we mean, outside the ergosphere which is located at
$r_{ergo}=M+\sqrt{M^2+n^2-a^2\cos^2\th}$.

\begin{figure}
  \begin{center}
\subfigure[along the pole]{
\includegraphics[width=2.5in,angle=0]{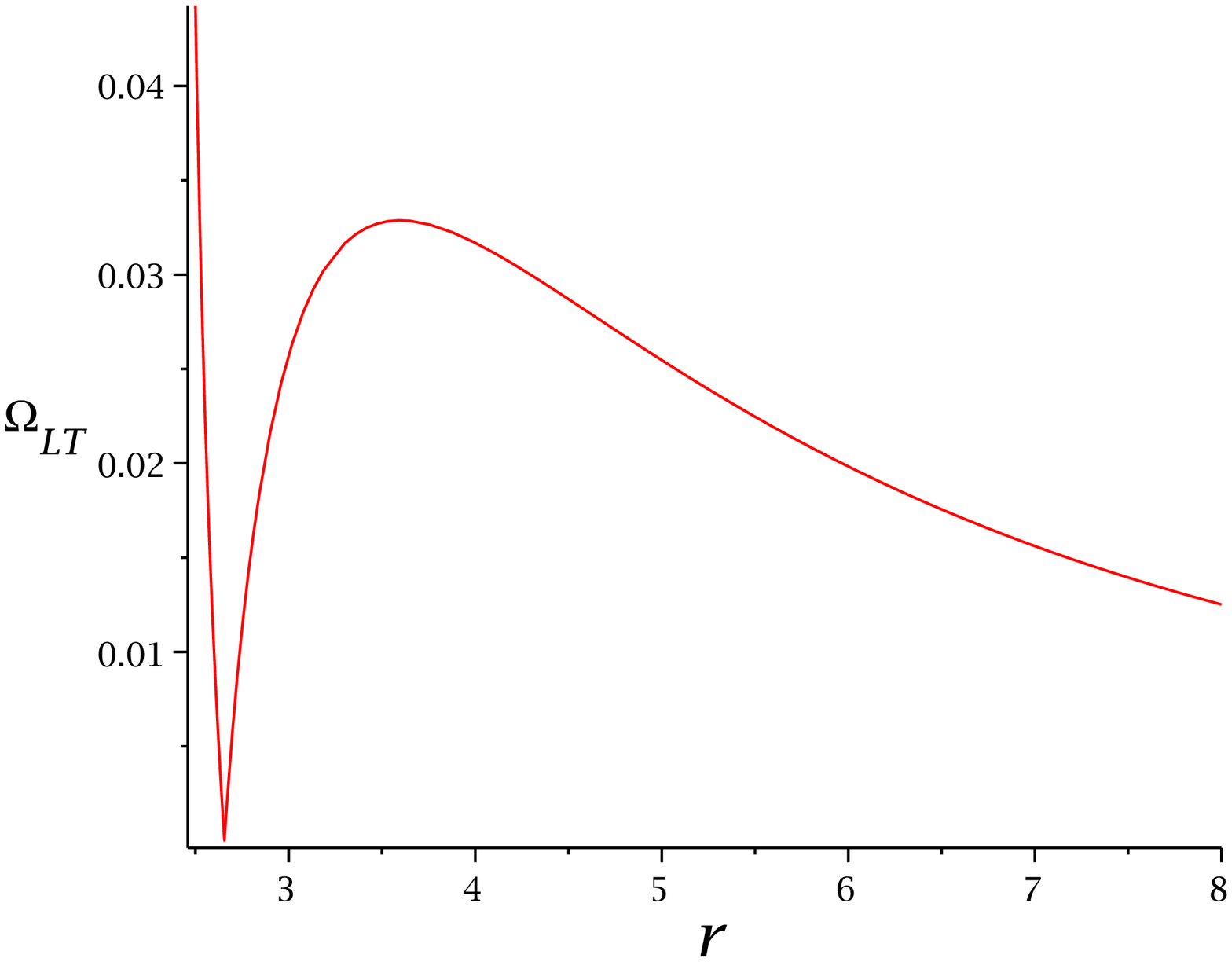}} 
\subfigure[along the equator]{
 \includegraphics[width=2.5in,angle=0]{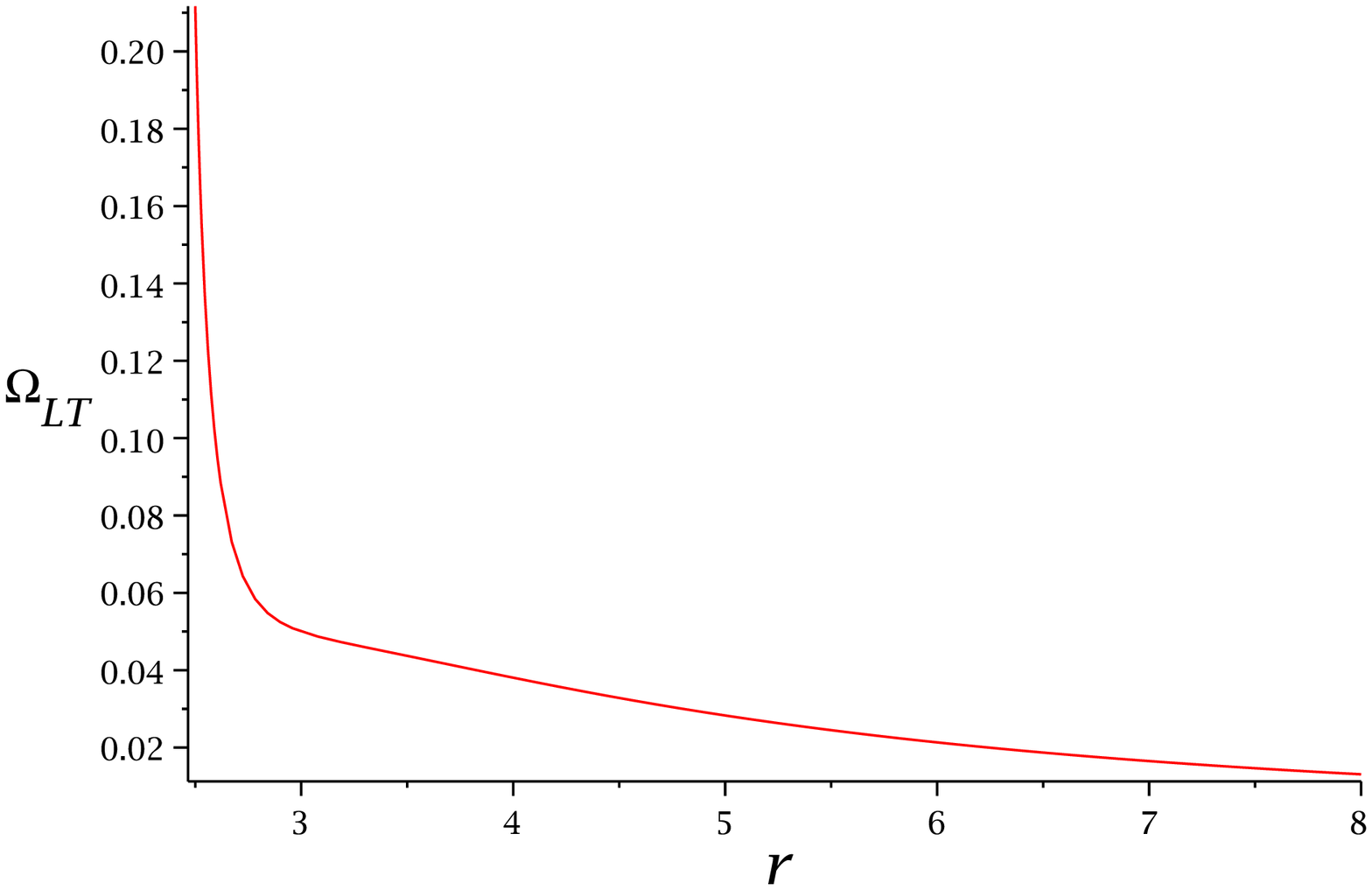}}
\caption{\label{a.1n1}\textit{Plot of $\O_{LT}$ vs $r$ 
in the KTN spacetime for $a=0.1\,\, m$,  $n=1\,\, m$ \& $M=1\,\, m$}}
% \label{a.1n1}
\end{center}
\end{figure}

\begin{figure}
  \begin{center}
\subfigure[along the pole]{
\includegraphics[width=2.5in,angle=0]{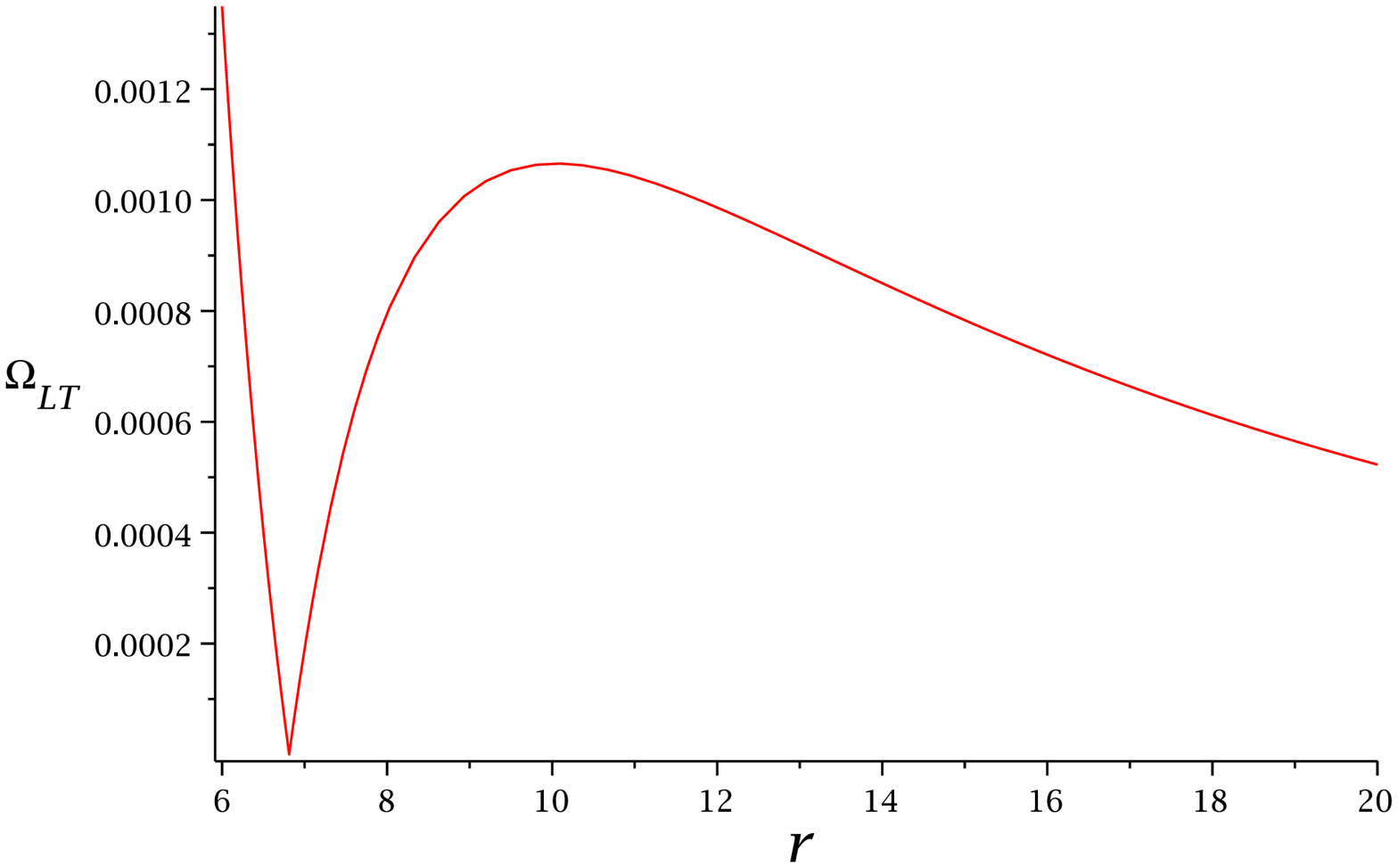}} 
\subfigure[along the equator]{
 \includegraphics[width=2.5in,angle=0]{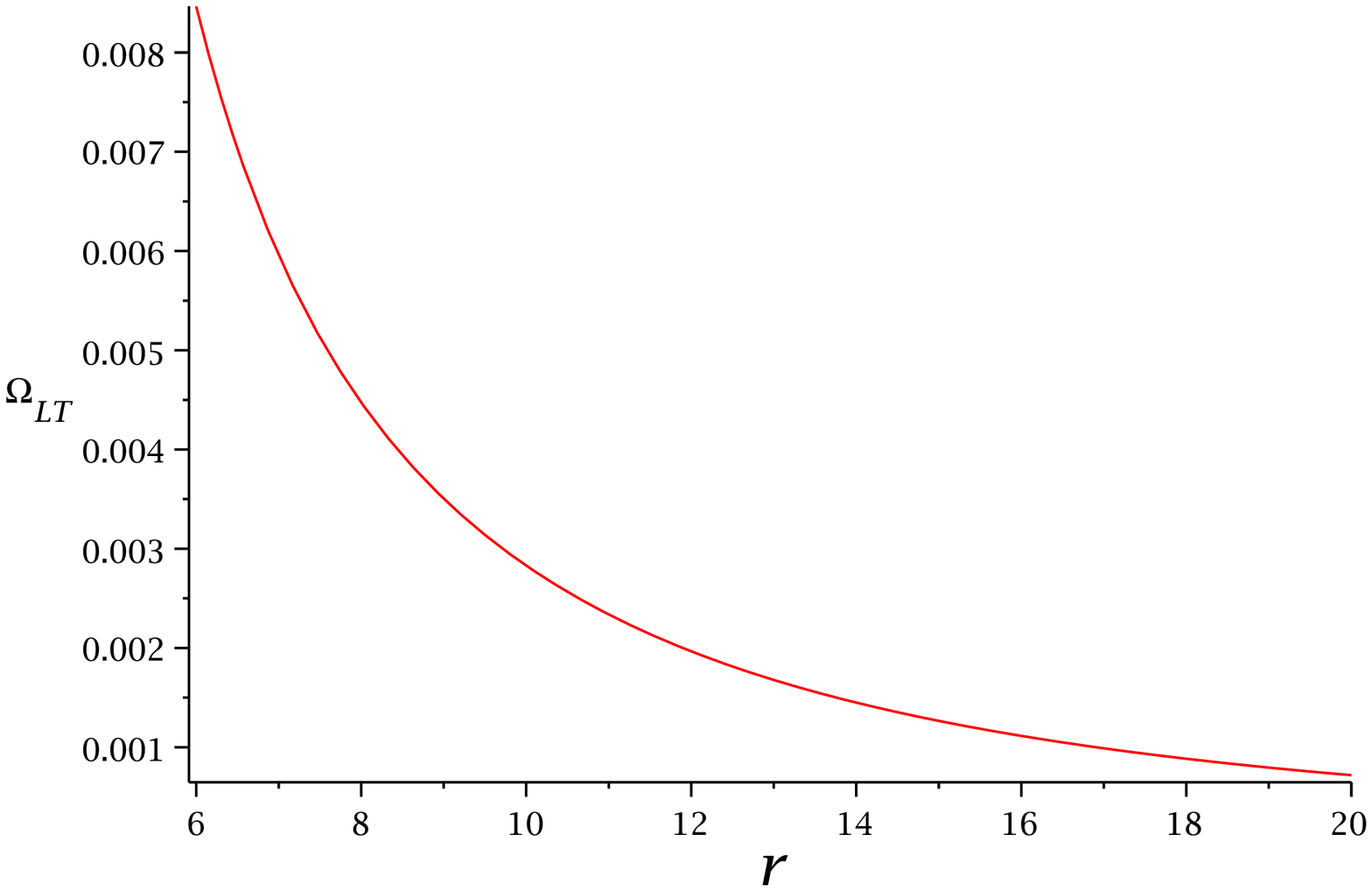}}
\caption{\label{a.7n.3}\textit{Plot of $\O_{LT}$ vs $r$ 
in the KTN spacetime for $a=0.7\,\, m$,  $n=0.3\,\, m$ \& $M=1\,\, m$}}
\end{center}
% \label{a.7n.3}
\end{figure}

We plot $r$ vs $\O_{LT}$ for $a<n$ (Fig. \ref{a.1n1}) 
and $a>n$ (Fig. \ref{a.7n.3})
 we see that the LT precession rate curve is smooth 
along the equator (panel (b)) but it is not smooth along the pole (panel (a)).
The LT precession rate along the pole is 
very high just outside the ergosphere and falls sharply
and becomes zero,
rises again and finally approaches to a small value after
crossing the very strong gravity regime. 
We will now discuss an interesting situation in which
the Kerr parameter $a$ is equal to the NUT parameter $n$.
\\

{\bf Special case $a=n$:}
The horizons of the KTN spacetime are located at $r_{\pm}=M\pm\sqrt{M^2+n^2-a^2}$
\cite{cc}. One horizon is located at $r_+>0$ and another 
is located at $r_-<0$ (if $n>a$)\cite{kag}.
  The Kerr parameter $a$ takes any value
but less than or equal to $\sqrt{M^2+n^2}$ in case of the KTN spacetime whereas 
$a$ takes its highest value as $M$ in case of the Kerr spacetime. 
Without this restriction (if $a^2>M^2+n^2$) the both spacetimes
lead to show the naked singularities.   
There are two special cases in KTN spacetimes for which $a$
can take the value $M$ only and for the second case $a$ can
take the value $n$. For the first case the angular momentum 
of the KTN spacetime would be $J=M^2$ which is similar to 
the case of extremal Kerr spacetime. In this case the horizons will be located 
at the distances $r_+=M+n$ and $r_-=M-n$. If the mass of the spacetime 
is greater than the dual mass of the spacetime ($M>n$), the both horizons could be
located at the positive distances $(r_{\pm}>0)$ but if the dual mass 
is greater than the mass of the spacetime ($M<n$) $r_-$ will be located
at the negative distance $(r_-<0)$.

For the second case $(a=n)$ the angular momentum of the 
KTN spacetime would be $J=Mn$. It is a very interesting 
situation. In this case the line element of the KTN spacetime 
would be 
\begin{equation}
ds^2_n=-\f{\d_n}{p_n^2}(dt-A_n d\phi)^2+\f{p_n^2}{\d_n}dr^2+p_n^2 d\th^2
+\f{1}{p_n^2}\sin^2\th(ndt-B_nd\phi)^2
\label{lnelmnt}
\end{equation}
with 
\begin{eqnarray}\nonumber
\d_n&=&r(r-2M),  p_n^2=r^2+n^2(1+\cos\th)^2,
\\
A_n&=& n(\sin^2\th-2\cos\th), B_n=r^2+2n^2.
\end{eqnarray}
It could be easily seen that this special rotating spacetime 
has outer horizon at the distance $r_+=2M$ 
and inner horizon at $r_-=0$. Outer horizon at the distance $r_+=2M$  
is just similar toversion accepted for publication in
the Schwarzschild spacetime where the event horizon is located at
$r=2M$. This spacetime can be treated as the rotating
spacetime with the {\it event horizon} at $r=2M$ and
its angular momentum will be
\begin{equation}
 J=Mn
\end{equation}
 In other words
it could be said that {\it the KTN spacetime rotating with 
the angular momentum $J=Mn$, possessed an outer horizon at $r=2M$
and an inner horizon at $r=0$.}
There is an apparent similarity between Eq. (40) of Ref. \cite{bh}
with our results but it is completely different situation.
Furthermore, there should be an
ergoregion in this special KTN spacetime. For this special case $(a=n)$,
the radius of the ergosphere for the KTN spacetime
will be $M+\sqrt{M^2+n^2\sin^2\th}$. The LT
precession rate in this special spacetime will be
\begin{eqnarray}
\O_{LT}|_{a=n}=\f{n}{p_n}\left[\d_n\left(\f{\cos\th}{r^2-2Mr-n^2\sin^2\th}
-\f{1+\cos\th}{p_n^2}\right)^2+\sin^2\th \left(\f{r-M}{r^2-2Mr-n^2\sin^2\th}
-\f{r}{p_n^2}\right)^2\right]^{\f{1}{2}}
\label{ktan}
\end{eqnarray}

\begin{figure}
  \begin{center}
\subfigure[along the pole]{
\includegraphics[width=2.5in,angle=0]{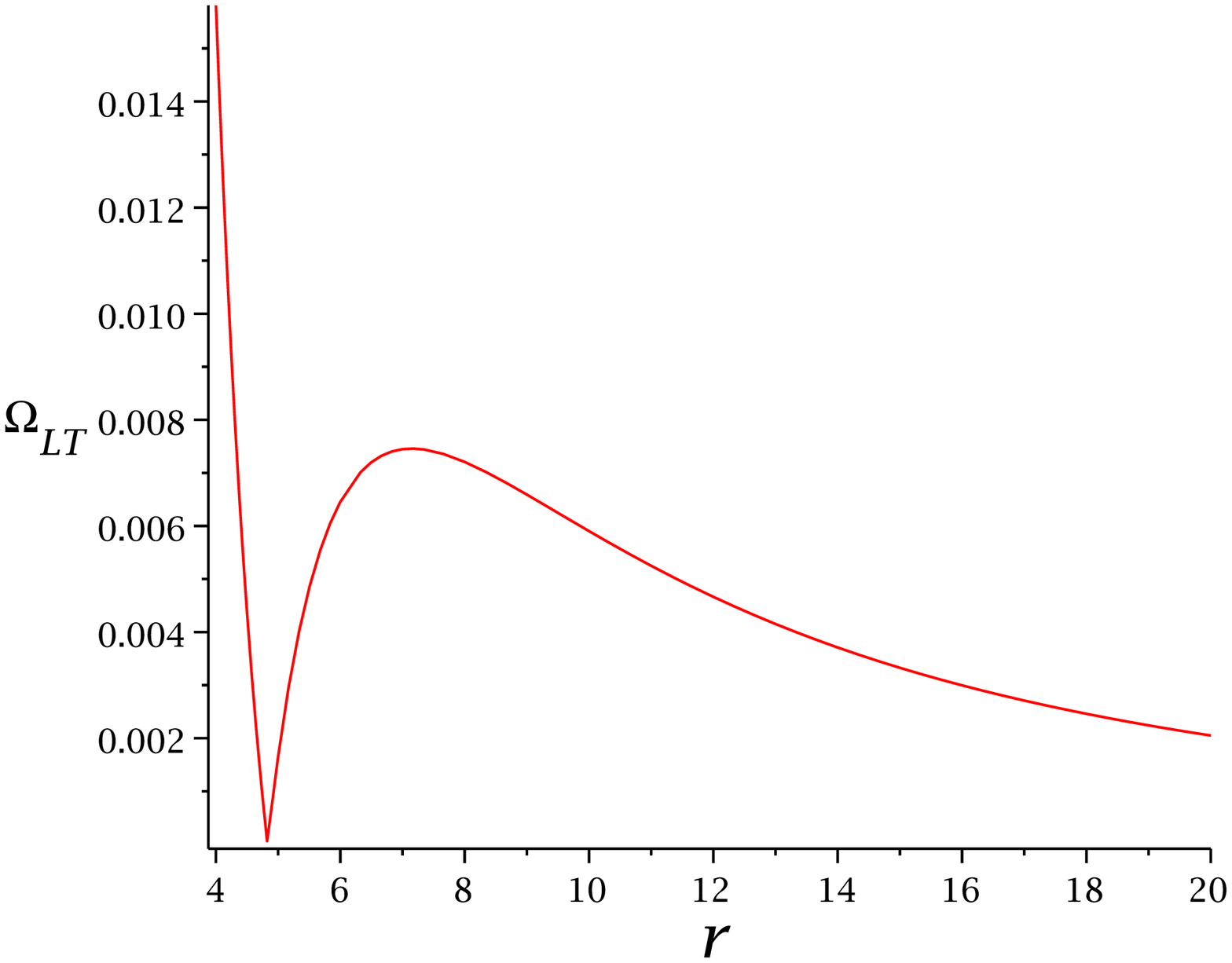}} 
\subfigure[along the equator]{
 \includegraphics[width=2.5in,angle=0]{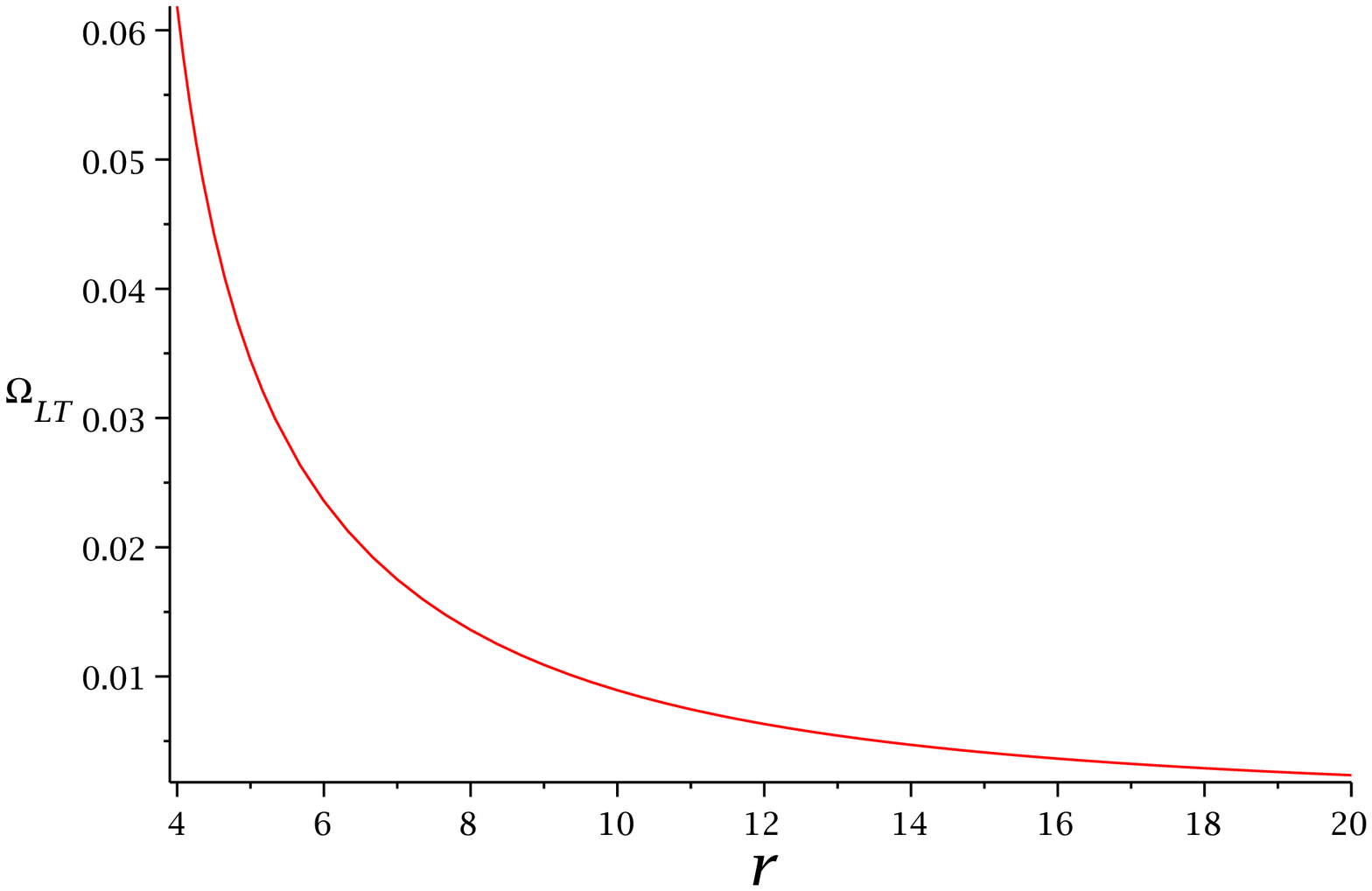}}
\caption{\label{an1}\textit{Plot of $\O_{LT}$ (in m$^{-1}$) vs $r$ (in m)
in the KTN spacetime for $a=n=1\,\, m$ \& $M=1\,\, m$}}
\end{center}
\end{figure}

The above expression is valid outside the ergosphere as it
diverges on the ergosphere and we also know that the LT
precession is not defined in the spacelike surface.
In Fig. \ref{an1}, we plot $r$ vs $\O_{LT}$ for $a=n=1$.
We see that the curve falls smoothly with increasing distance along the equator
but it is not smooth along the pole. Similar behaviour
is noticed in Fig. \ref{a.1n1} and Fig. \ref{a.7n.3}. 
The LT precession rate along the pole is 
very high just outside the ergosphere and falls sharply
and becomes zero,
rises again and finally approaches to a small value after
crossing the strong gravity regime. This is really 
very peculiar and this was not observed in any other spacetimes
which are the vacuum solutions of Einstein equation, previously.

\section{Results}

\begin{figure}
  \begin{center}
\subfigure[along the pole]{
\includegraphics[width=2.5in,angle=0]{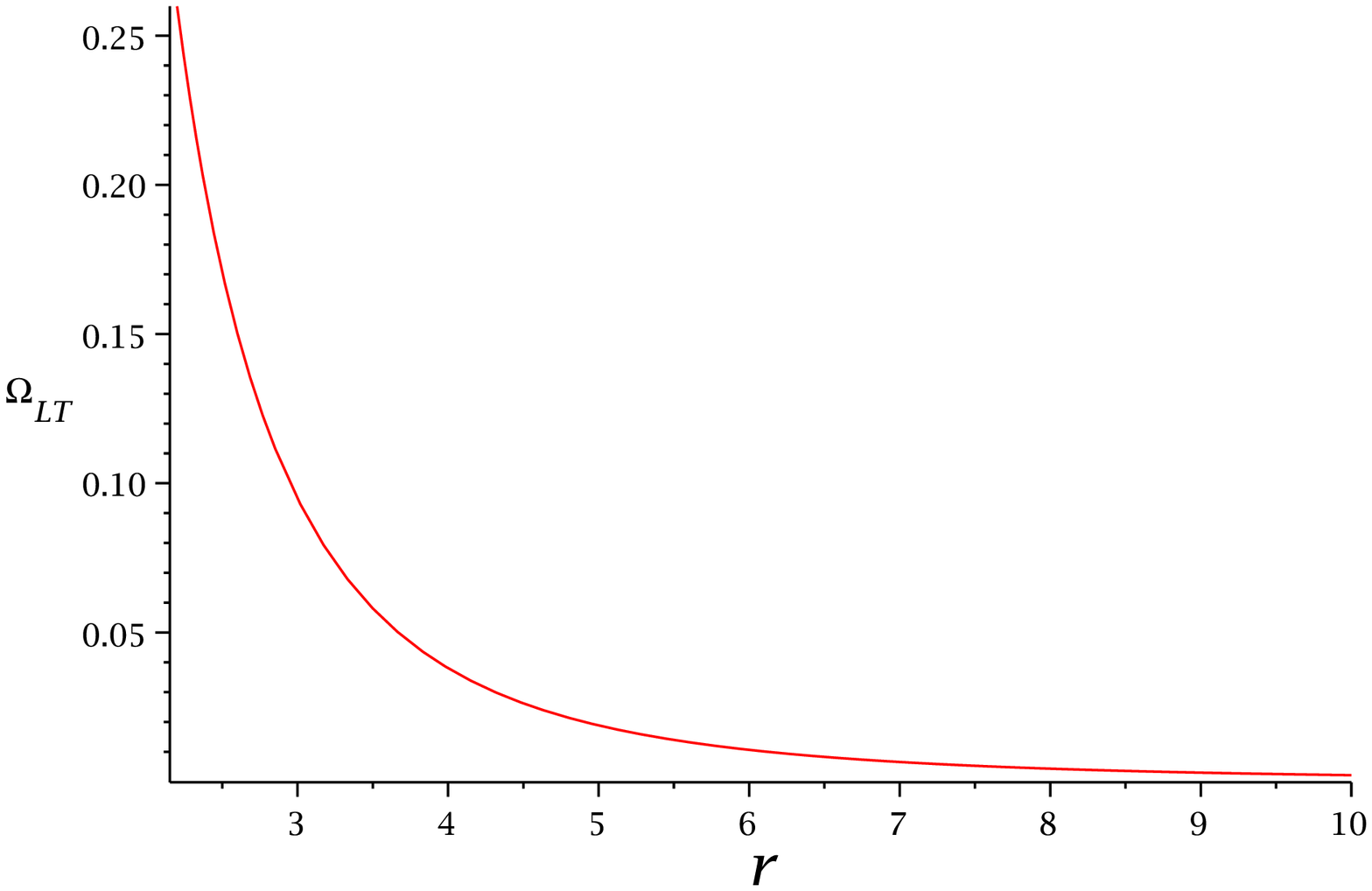}} 
\subfigure[along the equator]{
 \includegraphics[width=2.5in,angle=0]{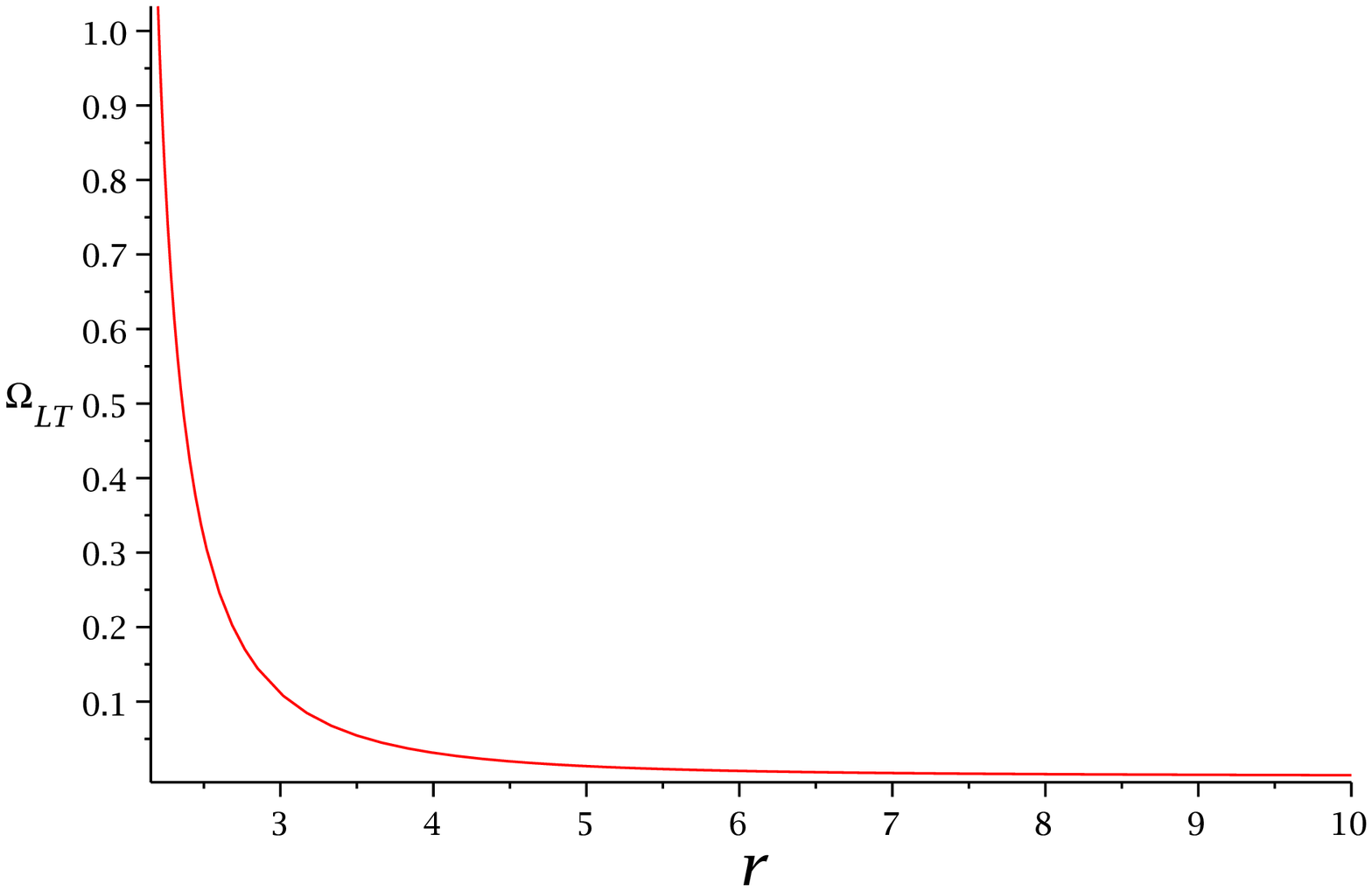}}
\caption{\label{a1}\textit{Plot of strong gravity $\O_{LT}$ (in m$^{-1}$) vs $r$ (in m)
in the Kerr spacetime for $a=M=1\,\, m$}}
\end{center}
\end{figure}

We know that the LT precession varies as $1/r^3$ in the weak
gravity regime (`weak' Kerr metric) by the famous relation
(Eq. 14.34 of Ref. \cite{jh})
\begin{equation}
\vec{\O}_{LT}=\f{1}{r^3}[3(\vec{J}.\hat{r})\hat{r}-\vec{J}] ~,
\label{we1}
\end{equation}
where, $\hat{r}$ is the unit vector along $r$ direction.
We plot $\O_{LT}$ vs $r$ in the strong gravity situation (see Eq. (42) 
of Ref. \cite{cm}) for maximally rotated Kerr spacetime along
the pole (panel(a)) and the equator (panel(b)) in Fig. \ref{a1}. 
Close observation
reveals that the LT precession rates at the same distances (for a fixed $r$) 
along the equator and the pole are not the same. In the strong gravity
regime $\O_{LT}^e$ is higher than $\O_{LT}^p$ as the ratio
$(\eta)$ of the LT precession rate along the pole $(\O_{LT}^p)$
to the equator $(\O_{LT}^e)$ in the strong gravity regime is
\begin{equation}
\eta_K^{strong}=\f{\O_{LT}^p}{\O_{LT}^e}
=\f{2r^3(r-2M)}{(r^2+a^2)^{\f{3}{2}}(r^2-2Mr+a^2)^{\f{1}{2}}} ~,
\label{rat_ks}
\end{equation}
but in the weak gravity regime it follows from Eq. (\ref{we1})
\begin{equation}
\eta_K^{weak}=\f{\O_{LT}^p}{\O_{LT}^e}=\f{\f{2J}{r^3}}{\f{J}{r^3}}=2 ~,
\label{rat_kw}
\end{equation}
which is a constant. If we look for this ratio in the case of
the KTN spacetime we find that 
 \begin{equation}
  \eta_{KTN}^{strong}=\f{\O_{LT}^p}{\O_{LT}^e}<1 ~.
\label{rat_kw}
 \end{equation}
It holds for ever i.e. the LT precession rate along the equator ($\O_{LT}^e$)
is always higher than the LT precession rate along the pole ($\O_{LT}^p$). 
In the weak gravity regime the ratio is only
 \begin{equation}
  \eta_{KTN}^{weak}=1 ~.
\label{rat_kw}
 \end{equation}
We can plot the ratio for the clear scenario.
\begin{figure}[h]
  \begin{center}
\subfigure[Kerr spacetime]{
\includegraphics[width=2.5in,angle=0]{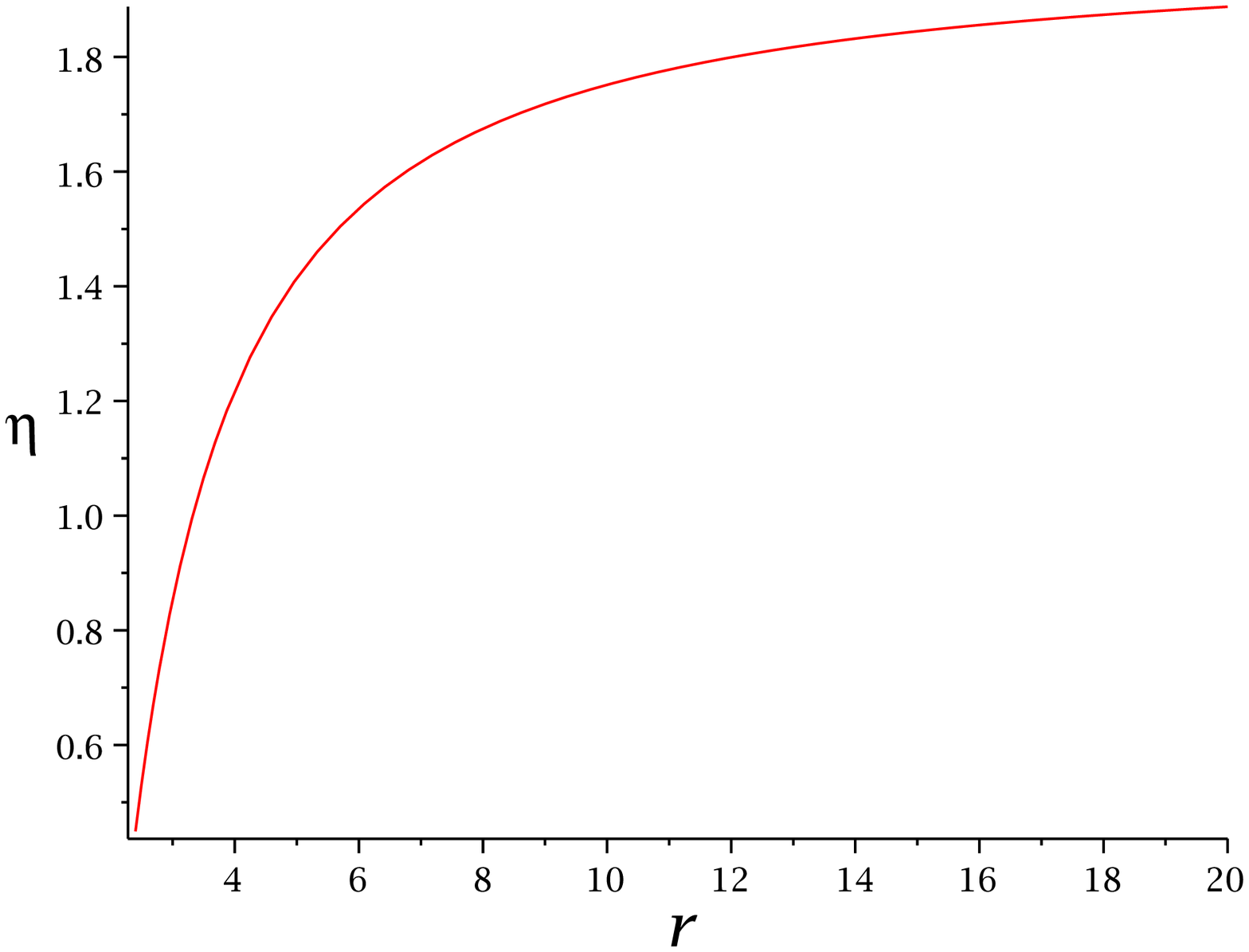}} 
\subfigure[KTN spacetime]{
 \includegraphics[width=2.5in,angle=0]{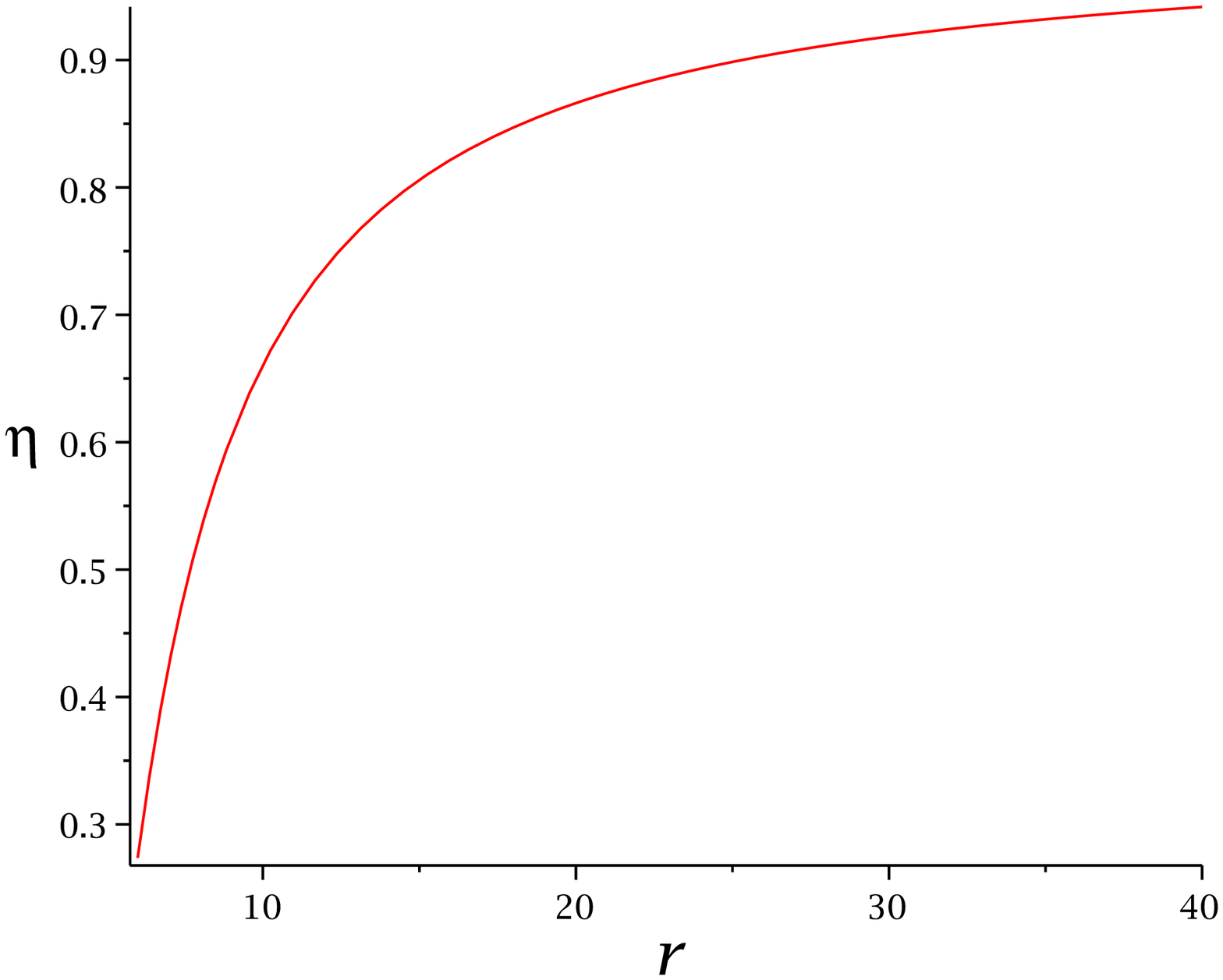}}
\caption{\label{rat}\textit{Plot of $\eta$ vs $r$ (in m)
in the Kerr and KTN spacetimes for $a=n=1\,\, m$ \& $M=1\,\, m$}}
\end{center}
\end{figure}
The plot in Fig. \ref{rat} for the Kerr spacetime shows that $\O_{LT}^p$ and $\O_{LT}^e$
are the same at a distance $r_0=3.324$ m. For $r<r_0$, $\O_{LT}^p<\O_{LT}^e$
and for $r>r_0$, $\O_{LT}^p>\O_{LT}^e$.

We have already seen that the plots of $\O_{LT}$ vs $r$ along the
pole and along the equator
both are smooth for the Kerr spacetime but this is not the same for 
the KTN spacetime. In the KTN spacetime though the curve of $\O_{LT}$ vs $r$ along the
the equator is smooth, it is not smooth along the pole. 
We have studied here basically three cases. These are following:
\\

\begin{figure}
    \begin{center}
\includegraphics[width=3in]{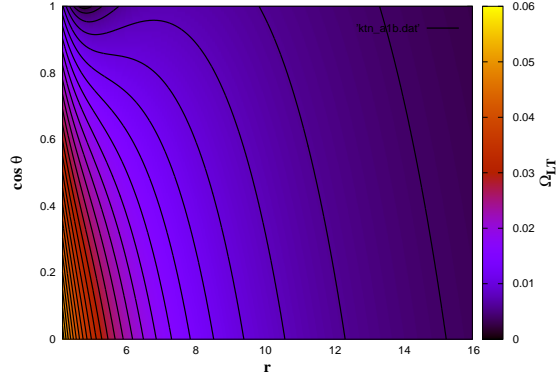}
      \caption{\label{an13}\textit{3-D plot of $\O_{LT}(r,\th)$
in the KTN spacetime for $a=n=1\,\, m$ \& $M=1\,\, m$}}
      \end{center}
\end{figure}

{\bf{(i) $a=n$ :}} In this case shown in Fig. \ref{an1}, we take the Kerr parameter $a$
is equal to the NUT parameter $n$ $(a=n=1$ m) and mass of the 
spacetime $M$ is unity. Thus, the radius of horizon 
is $r_+\sim2$ m. The LT precession rate along the pole (panel (a)) is tremendously high
just outside the horizon. Then it falls sharply and becomes zero (local minima)
at $r_{min}\sim4.8$ m. It rises again and gives a local maxima at  $r_{max}\sim7$ m.
After that the curve of the LT precession rate follows the
general inverse cube law and falls accordingly.
We cannot see the same feature along the equator. We plot a 3-D
picture of the LT precession rate in Fig. \ref{an13} 
where the $Y$ axis represents the cosine of colatitude ($\cos\th$)
and $X$ axis represents the distance $(r)$ from 
the centre of the spacetime. The colors 
represent the value of the LT precession rate and the 
values of the same precession rates are also separated by the isocurves.
It shows that there is a local maximum and local minimum along
the pole but it disappears after crossing a certain 
`critical' angle. Here, it is around $\cos\th\sim 0.6$.
\\

\begin{figure}
    \begin{center}
\includegraphics[width=3in]{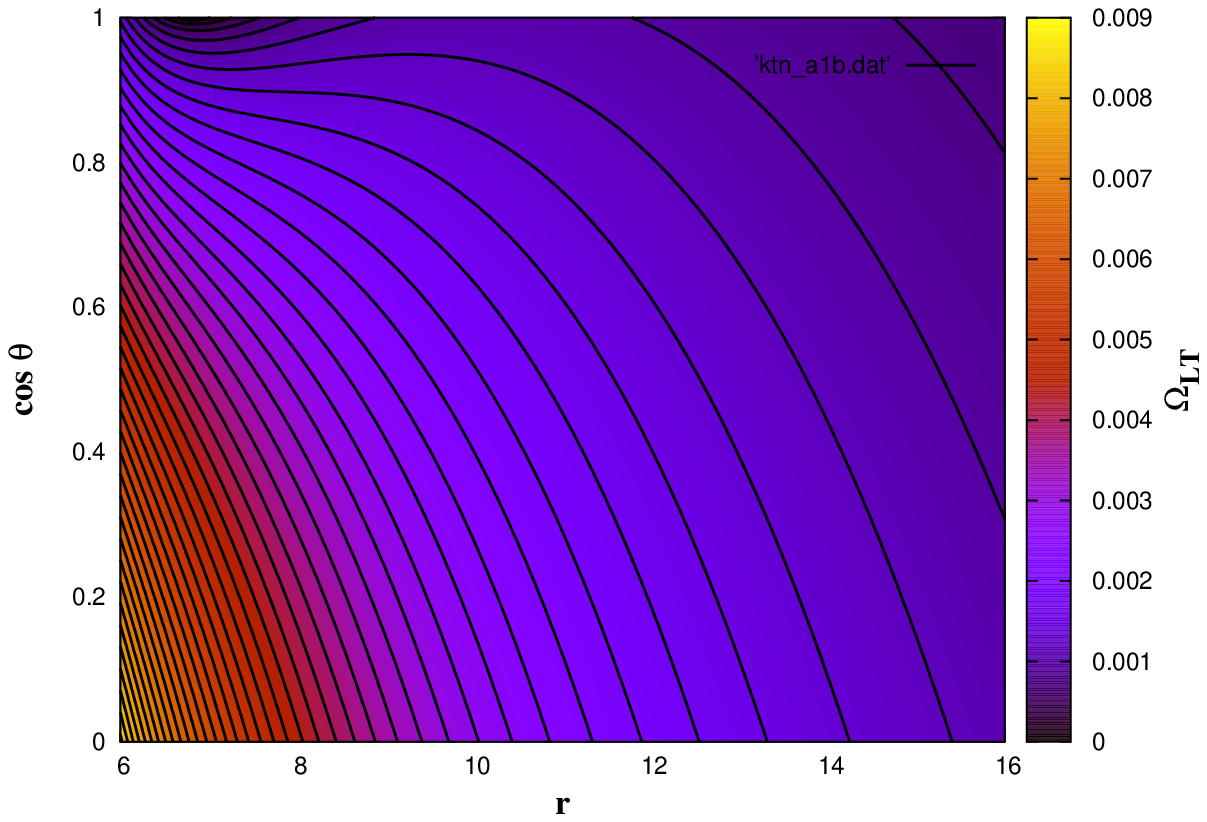}
      \caption{\label{a.7n.33}\textit{3-D plot of $\O_{LT}(r,\th)$
in the KTN spacetime for $a=0.7\,\, m$,  $n=0.3\,\, m$ \& $M=1\,\, m$}}
      \end{center}
\end{figure}

{\bf{(ii) $a>n$ :}} In the second case shown in 
Fig. \ref{a.7n.33}, the Kerr parameter $a=0.7$ m
and NUT parameter $n=0.3$ m. Mass of the spacetime is $M=1$ m. 
Radius of the horizon $r_+\sim1.8$ m, distance of local minimum is 
$r_{min}\sim7$ m and distance of local maximum is $r_{max}\sim10$ m .
The `critical' angle is around $\cos\th\sim 0.8$.
\\

\begin{figure}
    \begin{center}
\includegraphics[width=3in]{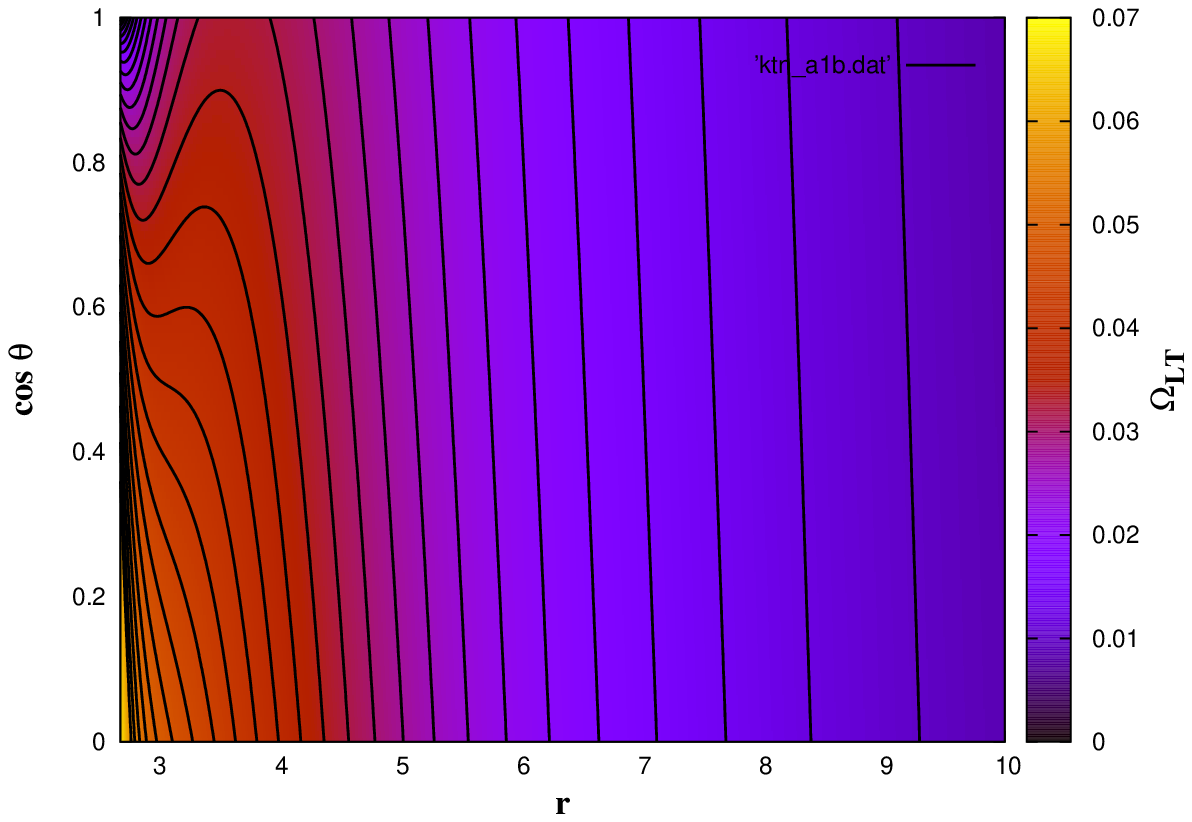}
      \caption{\label{a.1n13}\textit{3-D plot of $\O_{LT}(r,\th)$
in the KTN spacetime for $a=0.1\,\, m$,  $n=1\,\, m$ \& $M=1\,\, m$}}
      \end{center}
\end{figure}

{\bf{(iii) $a<n$ :}} For the third case exhibited in 
Fig. \ref{a.1n13}, the Kerr parameter $a=0.1$ m
and NUT parameter $n=1$ m. Mass of the spacetime is $M=1$ m. 
Radius of the horizon $r_+\sim2.4$ m, distance of local minimum is 
$r_{min}\sim 2.6$ m and distance of local maximum is $r_{max}\sim3.5$ m .
 The `critical' angle is around $\cos\th\sim 0.4$.
\\

\begin{figure}
    \begin{center}
\includegraphics[width=3in]{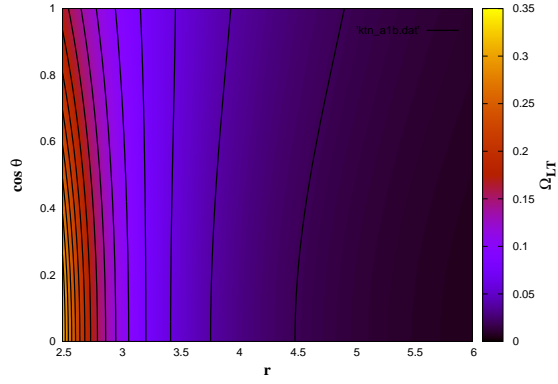}
      \caption{\label{a13}\textit{3-D plot of $\O_{LT}(r,\th)$ in the Kerr spacetime
 for $a=1\,\, m$ \& $M=1\,\, m$}}
      \end{center}
\end{figure}

In all three cases, plots show the same feature but 
the numerical values are different depending on the values of
$a$ and $n$. For a fixed value of $n$, if $a$ decreases
the value of the LT precession rate at the local maximum increases
and also the distance of local minimum and maximum are shifted 
towards the horizon of the spacetime. If the NUT parameter 
vanishes (for the Kerr spacetime) there will be no local
maximum and minimum as noticed in Fig. \ref{a1} and Fig. \ref{a13}.

\begin{figure}
  \begin{center}
\subfigure[plot of $\O_{LT}$ (in m$^{-1}$) vs $r$ (in m)]{
\includegraphics[width=2.5in]{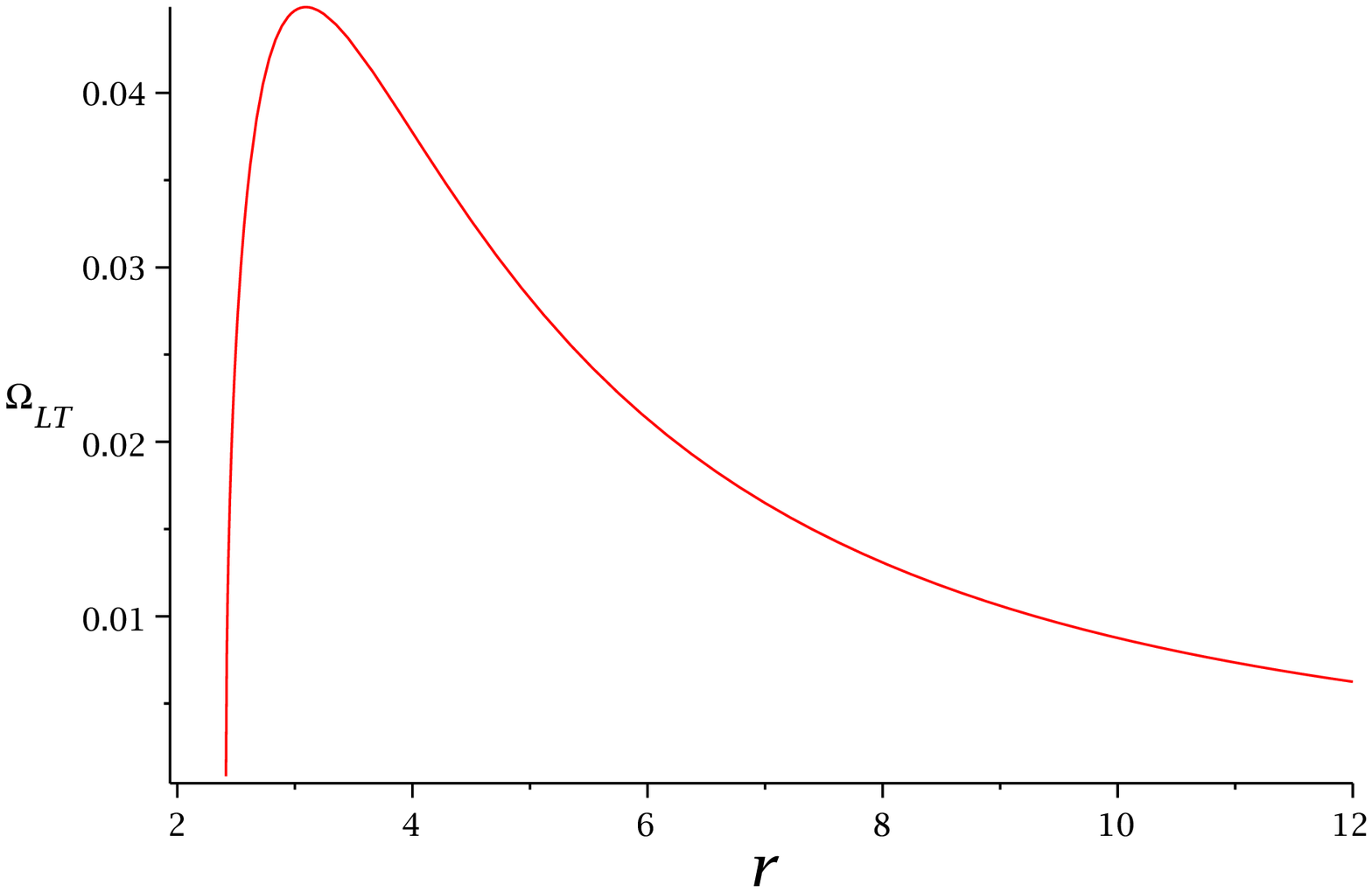}} 
\subfigure[3-D plot]{
 \includegraphics[width=2.5in]{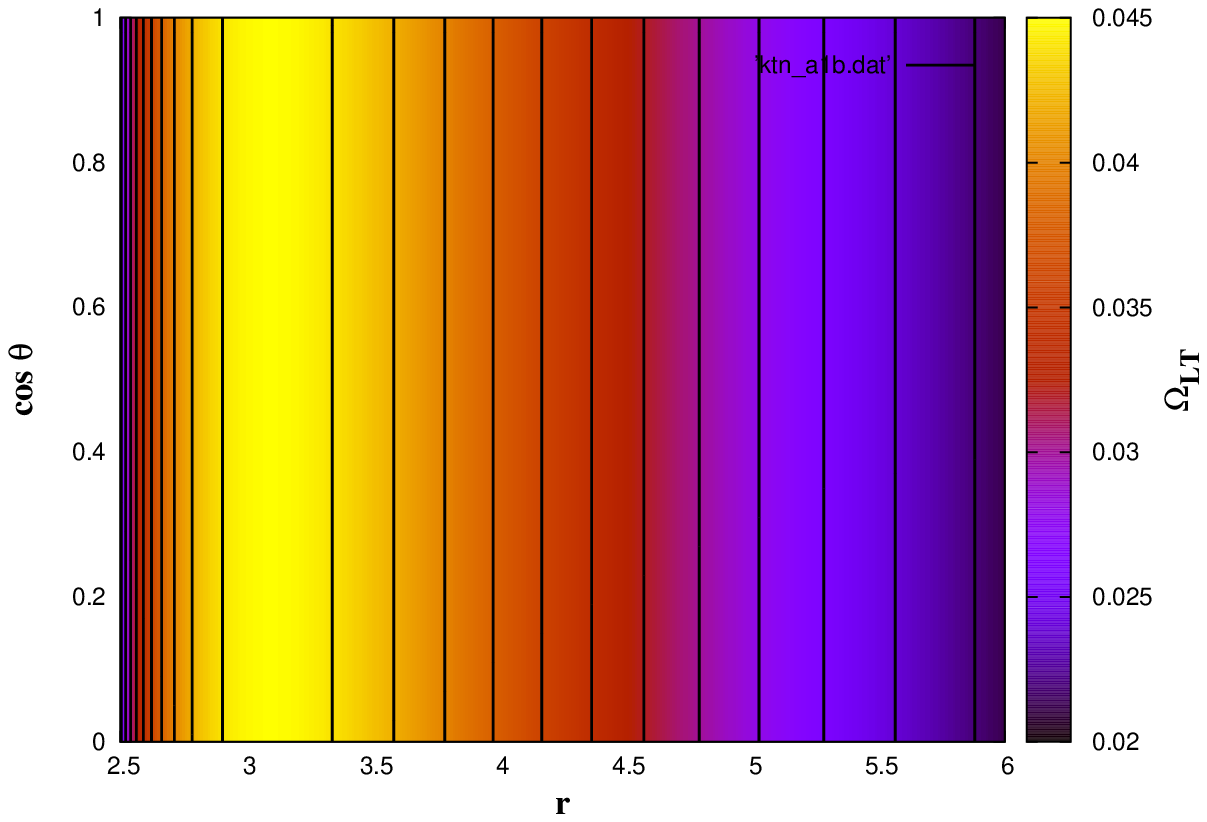}}
\caption{\label{n1}\textit{Plot of $\O_{LT}$ in the Taub-NUT spacetime
for $n=1\,\, m$ \& $M=1\,\, m$ 
(basically, the expression of
$\O_{LT}$ (see Eq. (25) of Ref. \cite{cm}) is independent of $\th$, 
 thus the value (colour) of $\O_{LT}$ does not change with $\cos\th$ in panel(b))}}
\end{center}
\end{figure}

The local maximum along the pole in 
the KTN spacetime arises due to the NUT parameter and
it is clear from Fig. \ref{n1} (panel(a)) that it is valid only for the
Taub-NUT spacetime where the Kerr parameter vanishes but the NUT parameter
does not vanish. The local minimum along the pole
in KTN spacetime arises due to the Kerr parameter.
It could not be seen directly from the Fig. \ref{a1}(panel(a)).
If we take Fig. \ref{a1}(panel(a)) of the Kerr spacetime and
Fig. \ref{n1}(panel(a)) of the Taub-NUT spacetime and overlap 
these two figures with each other (as the KTN spacetime includes
both the Kerr and NUT parameters) just for our clear 
understanding, we can easily
visualize the nature of the plots of the LT precession 
(panel(a) of Fig. \ref{a.1n1}-\ref{an1}) along the pole 
in the KTN spacetimes. Thus, the presence of the Kerr
parameter is responsible for showing the local minimum
along the pole in the KTN spacetime.

Without the Kerr parameter the LT precession rate
at a `local maximum' in the Taub-NUT spacetime is higher than 
the LT precession rate at a `local maximum' in the KTN spacetime.  
The presence of the Kerr parameter (or increasing the value
of the Kerr parameter from $0$ to a finite number) 
shifts the `local maximum' and `local minimum'
away from the horizon and reduces the LT precession rate at the 
local maximum.

We note that the Taub-NUT spacetime is a stationary and spherically symmetric
spacetime and the expression of 
$\O_{LT}$ (see Eq. (25) of Ref. \cite{cm}) is also independent of $\th$
(the LT precession in the Taub-NUT spacetime has been discussed in detail 
in the next section 4.4). 
Thus the value (colour) of $\O_{LT}$ does not change with $\cos\th$.
It means that the LT precession rate is same everywhere 
in that spacetime for a fixed distance $r$ (no matter whether it is
pole or equator) and the LT precession rate curve always shows a 
`peak' as seen in panel(a) of Fig. \ref{n1} near the horizon.
But, if this Taub-NUT spacetime starts to rotate
with an angular momentum $J(=aM, a$ is the Kerr parameter), it turns out to
be the KTN spacetime. In this case, the LT precession rate curve shows a 
`peak' (or `local maximum') along the pole but disappears after
crossing the `critical' angle and we cannot see any `peak' in the 
LT precession rate curve along the equator as discussed earlier. The `intrinsic'
angular momentum of the spacetime ($J$) is fully responsible for the no-show 
of `local maximum' along the equator. The Kerr
parameter is also responsible for reducing the LT precession
rate at the `local maximum' which has already been discussed
in the previous paragraph. Thus, the `dual mass' or the `angular 
momentum monopole' $n$ is only 
responsible for the `anomaly' (appearance of local maximum and local minimum
in the LT precession rate) and the Kerr parameter or the rotation 
of the spacetime tries to reduce this `anomaly' as far as possible. The Kerr parameter
is fully successful to reduce this effect along the equator
but slowly it loses its power of reduction of this anomaly 
along the pole.

\subsection{Appearance of the local maximum
and local minimum: A comparison study of some well-known spacetimes}
If we take the derivative of Eq. (\ref{ktmod}) with 
respect to $r$ and plot $\f{d\O_{LT}}{dr}|_{(r=R,\th=\pi/2)}$ 
vs $r$ we cannot find any positive real root in the region
$r_{ergo}<r<\infty$. But the plot of $\f{d\O_{LT}}{dr}|_{(r=R,\th=0)}$
vs $r$ shows two positive real roots (which are basically 
local maximum $R_1=r_{max}$ and local minimum $R_2=r_{min}$) 
in the region $r_{ergo}<r<\infty$. It has a similarity
with the case of the frame-dragging effect
inside the rotating neutron stars (see Appendix of \cite{cmb}).

It is very important to mention here that the LT 
precession rate (Eq.\ref{ktmod}) reduces to 
\begin{equation}
 \O_{LT}|_{\th=0}=\f{-nr^2+2Mr(n+a)+n(n+a)^2}{(r^2-2Mr+a^2-n^2)^{\f{1}{2}}
 (r^2+(n+a)^2)^{\f{3}{2}}}
\end{equation}
along the pole and it vanishes at 
\begin{equation}
 R_2=\left(1+\f{a}{n}\right) \left(M+\sqrt{M^2+n^2}\right)
\label{rmin}
\end{equation}
or,
\begin{equation}
 \f{R_2}{r_{+TN}}=\left(1+\f{a}{n}\right)
\label{rmin1}
\end{equation}
We note that the `event horizon' of the Taub-NUT spacetime 
is located at $r_{+TN}=M+\sqrt{M^2+n^2}$.
It means when Kerr parameter is zero, the $R_2$ goes to 
on the horizon in the case of Taub-NUT spacetime. Thus, 
the local minimum of the Taub-NUT spacetime and
the `horizon' coincide at the same point (see panel (a) of Fig.\ref{n1}).
As the value of $a$ increases from $0$ to a finite number, the 
position of local minimum shifts in outward direction. Thus,
the rotation of the spacetime is responsible for shifting
the local maximum and local minimum. We can check whether
the $r_{min}$ and $r_{max}$ are always outside the 
horizon $(r_{+KTN})$ or not. We can take the difference
of $r_{min}$ and $r_{+KTN}$ in KTN spacetime:
\begin{eqnarray}
 r_{min}-r_{+KTN}=a/n(M+\sqrt{M^2+n^2})+\sqrt{M^2+n^2}.(1-\sqrt{1-a^2/(M^2+n^2)})
\end{eqnarray}
As $a^2\geq(M^2+n^2)$, the above relation reveals that 
$r_{min}$ is greater than $r_{+KTN}$
which means $r_{min}$ always lies outside the horizon. 
This
not only holds along the pole but also for all angles.
For $\th>0$, $r_{min}$ lies outside the ergoregion. As it is difficult
to calculate the position of the `local minimum' analytically 
for all values of $\th$, we have plotted 
these for $0\leq\th\leq\pi/2$ and obtained the values numerically 
for few cases which has been described in the Results section.
For the extremal KTN spacetime,
\begin{eqnarray}
 r_{min}-r_{+KTN}=\sqrt{M^2+n^2}\left[1+\f{M}{n}+\sqrt{1+\f{M^2}{n^2}}\right] .
\end{eqnarray}

Kerr spacetime does not 
show up this type of anomaly (Fig. \ref{a1}). If we put $n=0$ in
Eq. (\ref{rmin}) for Kerr spacetime, $R_{2}$ does not make any sense.  
Is this same `anomaly' also appeared in the Kerr-Newman spacetime?
To get the answer, we can write the exact LT precession rate 
in Kerr-Newman spacetime \cite{cp}:
\begin{eqnarray}
\vec{\O}^{KN}_{LT}=\f{a}{\rho^3(\rho^2-2Mr+Q^2)}
\left[\sqrt{\d}(2Mr-Q^2) \cos\th \hat{r}
+(M(2r^2-\rho^2)+rQ^2)\sin\th \hat{\th}\right] .
\label{kn}
\end{eqnarray}
In the Kerr-Newman spacetime,
\begin{eqnarray}
 \d=r^2-2Mr+a^2+Q^2  \,\,\,\,
\text{and}\,\,\,\,
\rho^2=r^2+a^2 \cos^2\th 
\end{eqnarray}
wher $Q$ is the total charge of the spacetime.
The LT precession rate at the pole in Kerr-Newman spacetime is
\begin{eqnarray}
 |\O^{KN}_{LT}|_{\th=0}=\f{a(2Mr-Q^2)}{(r^2+a^2)^{\f{3}{2}}(r^2-2Mr+a^2+Q^2 )^{\f{1}{2}}} .
\end{eqnarray}
LT precession rate could vanish at $r=Q^2/2M$ and we may think that this 
particular point leads to a local minimum in Kerr-Newman spacetime as like as KTN spacetime.
To check this, we have to do a short calculation. Firstly, we take:
\begin{eqnarray}
 r_{min}=\f{Q^2}{2M}
\end{eqnarray}
and radius of the horizon:
\begin{eqnarray}
 r_{+KN}=M+\sqrt{M^2-Q^2-a^2}
\end{eqnarray}
We know that
\begin{eqnarray}
a^2 & \leq & M^2-Q^2 , 
\\
\f{a^2}{2M} & \leq & \f{M}{2}-\f{Q^2}{2M}.
\end{eqnarray}
Thus, 
\begin{eqnarray}
r_{min} & \leq &  M/2-a^2/2M ,
\\
r_{min}-r_{+KN} & \leq & -1/2M[(M^2+a^2)+2M(M^2-Q^2-a^2)^{1/2}] .
\label{knm}
\end{eqnarray}
Eq.(\ref{knm}) reveals that $r_{min}$ is always inside the 
horizon ($r_{+KN}$) of Kerr-Newman spacetime. This means that $r_{min}$
is in spacelike region where our basic formalism (Eq. (17) of \cite{cm})  
of LT precession
is not valid (as we know that the formalism is valid only in 
timelike surfaces, it has also been stated in previous sections) 
and to obtain the LT precession rate in this 
region is meaningless. Thus, $r_{min}$ of Kerr-Newman spacetime does not make any sense
and LT precession do not show any `anomaly' in this spacetime.
Now, for extremal Kerr-Newman spacetime Eq. (\ref{knm}) reduces to
\begin{eqnarray}
 r_{min}-r_{+KN} & \leq & -1/2M(2M^2-Q^2)
\end{eqnarray}
As $M>Q$, $r_{min}$ in extremal Kerr-Newman spacetime is also inside the horizon
and does not make sense. Thus, we can safely say that the `anomalous'
LT precession is absent in the Kerr-Newman spacetime. 

We can reiterate here that 
we derived the exact frame-dragging rate inside the rotating neutron
stars without making any assumption on the metric components as 
well as the energy-momentum tensor \cite{cmb}. We discussed our results for two types of pulsars:
(i) which rotate with their Kepler frequencies, and (ii) which rotate
with a frequency lower than their Kepler frequencies.
In the second case, we calculated the LT precession
frequencies for three real pulsars: J1807-2500B, J0737-3039A
and B1257+12. In both the cases, 
it was shown that the frame-dragging rate 
monotonically decreases from the centre to the 
surface of the neutron star along the pole. In case of 
frame-dragging
rate along the equatorial distance, it decreased initially away from the 
centre, became negligibly small well before  the surface of the neutron
star, rose 
again and finally approached to a small value at the surface \cite{cmb}. The appearance
of local maximum and minimum in this case was the result of the dependence of 
frame-dragging frequency on the distance and angle. 
Moving from the equator to the pole, it was observed that this local maximum and
minimum in the frame-dragging rate along the equator
disappeared after crossing a critical angle. 
It was noted that the positions 
of local maximum and minimum of the frame-dragging rate along the
equator depend on the rotation frequency and central energy density of a 
particular pulsar. We had also estimated the LT precession 
frequencies at the centers of these pulsars without imposing
any boundary conditions on them. The whole prescription
revealed that the LT precession rate in strong gravity regime depended not only on the 
distance $r$ but also on the colatitude $\th$ of the gyroscope.

After the above discussion we can say that the local maxima and 
local minima are appearing only in the KTN spacetimes
and the interior spacetimes of the rotating neutron stars. 
There may exist an indistinct relation between these two spacetimes.
We discuss it in the next section. All other spacetimes 
which we have checked, do not show this type of anomaly and it could be
said easily that this anomalous LT precession in the KTN 
spacetime is appearing only due to the NUT charge $(n)$.

\section{Summary and discussion}
We have shown that the LT precession in KTN and Taub-NUT spacetimes are quite different
than the LT precession in other spacetimes. Other vacuum solutions
of Einstein equation do not show this type of strange 
feature in the LT precession or frame-dragging effect. 
It has been discussed that this strangeness in the KTN spacetime is due to 
only the presence of the NUT parameter or (gravito)magnetic monopoles.
Remarkably, it (frame-dragging effect 
in the KTN spacetime) has an apparent similarity with
the frame-dragging effect inside the rotating neutron star. Exact
frame-dragging effect inside the rotating neutron star has recently
been derived and discussed in detail by Modak, Bandyopadhyay and myself
\cite{cmb} but this is the interior solution of the Einstein equation,
not the vacuum solution. In the case of the interior of
a pulsar, the LT precession shows the same `anomaly' like the KTN
spacetime but there is also an another basic difference:
the anomaly appears in the LT precession rate in the KTN spacetime along the 
pole but it appears along the equator in the case of a pulsar.
The basic features of the plots are same for both the cases.
We do not know if there is any connection or not in these two
spacetimes. 
 We have already stated that Lynden-Bell and Nouri-Zonoz \cite{lnbl}
first highlighted about the observational possibilities for NUT charges
or (gravito)magnetic monopoles and they claimed
that the signatures of such spacetime might be
found in the spectra of supernovae, quasars, or active 
galactic nuclei. In our case, the resemblance of the anomalous LT precessions 
in the KTN spacetimes and the spacetime of the pulsars could be the indirect proof of the existence of 
the NUT charge ((gravito)magnetic monopoles) inside the pulsars and
we also suggest that such a signature could be used to identify a role of Taub-NUT solutions
in the astrophysical observations. We hope that we shall be able to give a direct mathematical proof of 
the existence of (gravito)magnetic monopoles inside the pulsars in a future publication \cite{cc3}.
\\

\textbf{Acknowledgments :} 
Thanks due to K. P. Modak and G. Das for valuable suggestions.
I am grateful to the Department of Atomic Energy
(DAE, Govt. of India) for the financial assistance.

\end{document}